\documentclass[sigplan,nonacm]{acmart}

\usepackage{booktabs}
\usepackage{tabularx}
\newcommand{\cmark}{\textcolor{green!55!black}{\checkmark}}
\usepackage{tcolorbox}
\usepackage{float}
\usepackage{placeins} 

\usepackage{subcaption}
\usepackage{listings}
\usepackage{amsmath}
\usepackage{subcaption}
\usepackage{enumitem}
\usepackage[ruled,vlined]{algorithm2e}
\usepackage{balance}
\usepackage{adjustbox}

\usepackage{multirow}

\usepackage{algorithmic}

\definecolor{codegreen}{rgb}{0,0.6,0}
\definecolor{codegray}{rgb}{0.5,0.5,0.5}
\definecolor{codepurple}{rgb}{0.58,0,0.82}
\definecolor{backcolour}{rgb}{0.95,0.95,0.92}

\definecolor{oliveGreen}{RGB}{22, 139, 22}

\newcommand\seokjin[1]{{\textcolor{black}{#1}}}

\usepackage{textcomp}

\newcommand{\mithreehundred}{AMD Instinct\texttrademark MI300X\xspace}
\newcommand{\mithreetwentyfive}{AMD Instinct\texttrademark MI325X\xspace}

\newcommand{\ignore}[1]{}
\newcommand{\ours}[0]{ViBE\xspace}

\newcommand{\niparagraph}[1]{\vspace{0pt}\noindent\textbf{#1}}

\begin{document}

\title[ViBE: Co-Optimizing Workload Skew and Hardware Variability for MoE Serving]{ViBE: Co-Optimizing Workload Skew and \\ Hardware Variability for MoE Serving}




\author{Seokjin Go}
\affiliation{%
  \institution{Georgia Institute of Technology}
  \city{Atlanta}
  \state{GA}
  \country{USA}
}
\email{seokjin.go@gatech.edu}

\author{Marko Scrbak}
\affiliation{%
  \institution{Advanced Micro Devices, Inc.}
  \city{Austin}
  \state{TX}
  \country{USA}
}
\email{marko.scrbak@amd.com}

\author{Ephrem Wu}
\affiliation{%
  \institution{Advanced Micro Devices, Inc.}
  \city{Santa Clara}
  \state{CA}
  \country{USA}
}
\email{ephrem.wu@amd.com}

\author{Srilatha Manne}
\affiliation{%
  \institution{Advanced Micro Devices, Inc.}
  \city{Bellevue}
  \state{WA}
  \country{USA}
}
\email{srilatha.manne@amd.com}

\author{Divya Mahajan}
\affiliation{%
  \institution{Georgia Institute of Technology}
  \city{Atlanta}
  \state{GA}
  \country{USA}
}
\email{divya.mahajan@gatech.edu}

\begin{abstract}
In distributed Mixture-of-Experts (MoE) inference, input-dependent token routing interacts with GPU performance variability to create persistent stragglers under synchronized execution, where the slowest GPU determines layer latency.
This performance variability is inherent to modern accelerators: manufacturing variation, power limits, and thermal conditions introduce measurable execution-time differences across nominally identical GPUs.
The core challenge is that MoE execution-time imbalance arises from the interaction of \emph{workload skew} and \emph{hardware asymmetry}. Token routing produces uneven and layer-varying expert loads, while GPU throughput depends on device-specific operating characteristics and workload intensity.
Prior work mitigates routing skew but assumes homogeneous hardware, optimizing token balance rather than execution latency. As a result, even balanced token assignments can leave hardware-induced stragglers unaddressed.

Thus, we propose \textit{Variability-Informed Binning of Experts} (\ours), a hardware-aware expert placement framework that minimizes execution-time imbalance across GPUs.
\ours combines per-GPU performance modeling with expert activation profiling to assign high-load experts to faster devices and low-load experts to slower ones, reducing layer-level stragglers without modifying model semantics or hardware.
Because both workload characteristics and effective GPU throughput can shift across serving conditions, \ours supports lightweight recalibration under workload/performance drift to refresh its routing and performance estimates when needed. This preserves placement accuracy across changes in token magnitude, batch composition, and execution phase while avoiding unnecessary reshuffling overhead.
Results show that \ours consistently reduces execution-time imbalance and improves SLO attainment by \seokjin{14\%}, while lowering P90 TTFT by up to \seokjin{45\%}. We further show that the impact of hardware variability increases at scale, making variability-aware placement important for efficient, high-utilization LLM serving.

\end{abstract}
\maketitle
\section{Introduction}

\begin{figure}[t]
\centering
\includegraphics[width=0.8\linewidth]{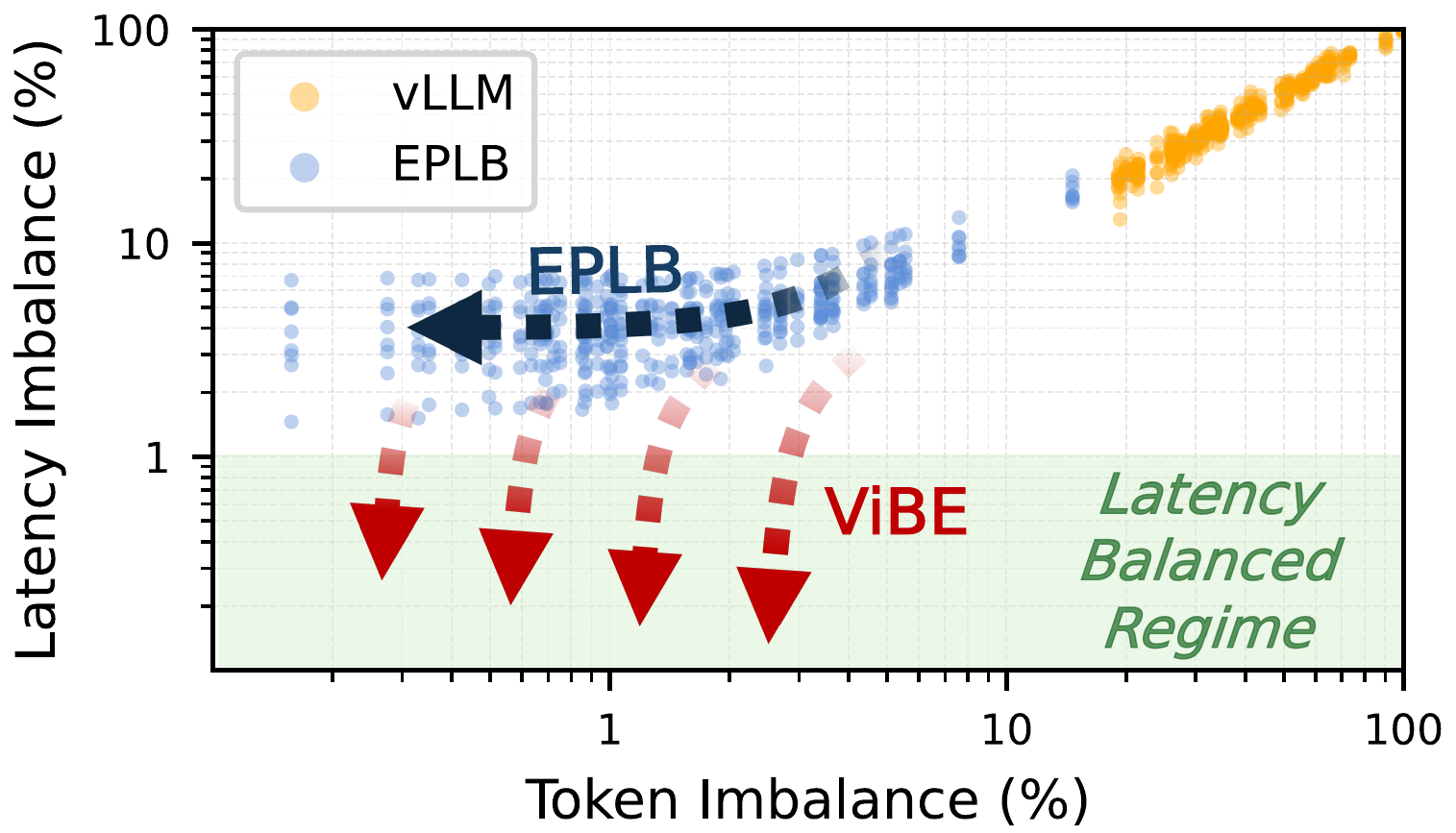}
\vspace{-0.5em}
\caption{Token vs.\ latency imbalance across MoE layers of DeepSeek-V3.
Each point is one layer; axes show max/min ratio within a layer.
EPLB reduces token imbalance but latency imbalance persists, while the proposed work directly targets the latency-balanced regime.}
\vspace{-1em}

\label{fig:pareto}
\end{figure}

Mixture-of-Experts (MoE) architectures allow large language models to scale by decoupling total model capacity from per-token computation~\cite{deepseek,mixtral,llama4,switchtransformer,gshard,qwen,gptoss}. Instead of activating all parameters for every token, MoE models route each token to a small subset of experts, allowing parameter count to grow without a proportional increase in compute.

This sparsity introduces system-level challenges during distributed inference. Each expert's parameters must reside in device memory, requiring model components to be partitioned across multiple GPUs or nodes. While data parallelism (DP), tensor parallelism (TP), and pipeline parallelism (PP) are general mechanisms for scaling large models, MoE systems additionally rely on expert parallelism (EP) to place expert weights across GPUs since the full set of experts cannot fit within a single device’s memory~\cite{megatron,huang2019gpipe,narayanan2019pipedream,switchtransformer,deepspeed-moe,phaze,nest}.
With EP, each MoE layer requires all-to-all communication and synchronization across GPUs and layer latency is determined by the GPU that finishes last. This bottleneck can arise from routing skew, hardware performance variability, or their interaction. Routing skew is a challenge in MoE inference because a small subset of experts can receive a large fraction of tokens, creating hotspots that increase execution time on some GPUs while others remain underutilized~\cite{eplb,megascaleinfer,moetuner,fastermoe}.

The key challenge in distributed MoE inference is persistent execution-time imbalance across GPUs caused by the interaction of routing-induced workload skew and hardware performance variability. Token routing creates uneven and layer-varying expert loads, while even nominally identical GPUs, across both AMD and NVIDIA platforms, can differ in effective throughput due to manufacturing variation, power limits, and operating conditions~\cite{notallgpus,pal,charllm}. Prior work reports up to 22\% variation in kernel execution time across compute-bound GPUs and shows that this variability is widespread across devices and nodes rather than confined to a few anomalous systems~\cite{notallgpus,gpulottery}. Under EP, these differences accumulate because each layer synchronizes at the pace of the slowest GPU, reducing throughput and increasing tail latency~\cite{fastermoe,li2023accelerating,comet}. At scale, this tail latency translates directly into lower effective utilization and higher serving cost since operators pay for provisioned GPU capacity that cannot be converted into useful inference throughput.

Past works have addressed these issues in isolation. Algorithmic approaches re-balance experts to equalize token counts, assuming homogeneous hardware~\cite{eplb,moetuner}. Hardware-level techniques address variability through power-aware scheduling or throughput-oriented management~\cite{pal,throttllem,perseus,tapas,dynamollm,litsilicon}. However, as \autoref{fig:pareto} illustrates, addressing only one source of imbalance is insufficient: EPLB successfully reduces token imbalance across MoE layers (points shift left toward the $y$-axis), yet latency imbalance persists because GPUs within the same node can still differ by up to 7\% in MoE kernel time (Figure~\ref{fig:eplb_valb_kerneltime}). Equalizing tokens does not equalize completion times given the underlying hardware variability.

Existing dynamic rebalancing mechanisms are also limited in how they adapt. Prior approaches~\cite{eplb} periodically recalibrate placement based on token distribution, but they consider only the relative routing ratio across experts or GPUs, not the absolute token magnitude that determines workload stress and exposes hardware variability. As a result, they cannot capture changes in performance asymmetry caused by shifts in batch size or serving phase. In addition, recalibration is typically invoked at fixed intervals that is not workload-aware and can introduce reshuffling overhead even when the placement remains appropriate.

\textbf{\textit{Our key insight is that routing skew and hardware variability should be co-optimized rather than treated independently.}}
Instead of attempting to eliminate variability, we use it as a lever to balance execution time. By assigning high-load experts to faster GPUs and low-load experts to slower GPUs, we can balance execution time even when token counts are not equal. This transforms variability into an opportunity to reduce stragglers and tail latency.

To realize this insight, we propose Variability-Informed Binning of Experts (\ours), a hardware-aware expert placement framework that directly minimizes execution-time imbalance across GPUs.
\ours combines per-GPU performance modeling with expert activation profiling to guide placement. It assigns high-load experts to faster devices and low-load experts to slower devices, balancing execution time across EP ranks without modifying model semantics or hardware.
As workload characteristics and effective GPU throughput can shift during serving, \ours also supports lightweight recalibration under workload/performance drift. Rather than recalibrating at fixed intervals, \ours monitors batch token count and observed latency imbalance, and refreshes its routing and performance estimates only when drift exceeds a configurable threshold. This makes recalibration stress-aware and workload-aware while reducing unnecessary reshuffling overhead.
Unlike token-balanced approaches, \ours minimizes predicted layer latency span rather than token count alone, producing more consistent execution across GPUs.

We evaluate \ours across two representative MoE models and two GPU generations, \mithreetwentyfive and \mithreehundred.
Results show that \ours improves kernel-level balance, reduces TTFT and TPOT tail latency, and increases sustainable request throughput compared to token-based balancing methods~\cite{eplb}. We also observe that token-balanced placement can assign high-load experts to slower GPUs, leading to suboptimal performance when hardware variability is ignored.

Overall, this work makes the following contributions:

\begin{itemize}[leftmargin=*,nosep]

\item \textbf{Joint characterization of routing skew and hardware variability.}
We quantify how these factors interact to create persistent stragglers, observing up to 7\% kernel execution time variation even under balanced token loads.

\item \textbf{Co-optimization of workload skew and hardware asymmetry.}
We show that expert placement can use routing skew to offset device-level performance differences, shifting the objective from token balancing to execution-time balancing.

\item \textbf{Variability-Informed Binning of Experts (\ours).}
We introduce a hardware-aware placement strategy that minimizes execution-time imbalance by assigning high-load experts to faster GPUs and low-load experts to slower GPUs.

\item \textbf{Drift-aware recalibration.}
We augment placement with lightweight recalibration under workload/performance drift, using token magnitude and latency imbalance to trigger updates only when needed.

\item \textbf{Evaluation across models and platforms.}
We demonstrate consistent improvements in load balance, tail latency, and SLO attainment across multiple MoE models and GPU platforms.

\end{itemize}

\begin{table*}[t]
\caption{We categorize pertinent prior work across four variability sources; {\color{green!55!black}\checkmark} indicates that a source is explicitly addressed.
}
\label{tab:related-work}
\setlength{\tabcolsep}{5pt}
\renewcommand{\arraystretch}{1.2}
\resizebox{0.98\linewidth}{!}{
\footnotesize\begin{tabularx}{\linewidth}{%
  l
  >{\centering\arraybackslash}p{1.2cm}
  >{\centering\arraybackslash}p{1.2cm}
  >{\centering\arraybackslash}p{1.2cm}
  >{\centering\arraybackslash}p{1.2cm}
  X}
\toprule
\textbf{System} &
\textbf{Hardware\textsuperscript{*}} &
\textbf{Phase} &
\textbf{MoEs} &
\textbf{Service} &
\textbf{Key Mechanism} \\
\midrule
PAL~\cite{pal}                 & \cmark &        &        &        &
  Power-aware scheduling under heterogeneous GPU performance \\

TAPAS~\cite{tapas}             & \cmark &        &        &        &
  Thermal-aware placement with power capping \\

Perseus~\cite{perseus}             & \cmark &        &        &        &
  Pipeline scheduling and frequency capping for energy saving \\

DynamoLLM~\cite{dynamollm}     & \cmark &      &        & \cmark &
  DVFS tuning and dynamic parallelism for workload variability \\

Splitwise~\cite{splitwise}     &        & \cmark &        &        &
  Prefill/decode disaggregation across heterogeneous devices \\

Sarathi~\cite{sarathi}         &        & \cmark &        &        &
  Chunked prefill with continuous batching \\

AlpaServe~\cite{alpaserve}     &        &        &        & \cmark &
  Multi-instance scheduling for request-level variability \\

EPLB~\cite{eplb}               &        &        & \cmark &        &
  Routing-frequency-based expert placement \\

\midrule
\textbf{ViBE (Ours)}           & \cmark & \cmark & \cmark & \cmark &
  \textbf{Variability-informed expert placement to minimize latency imbalance} \\
\bottomrule
\multicolumn{6}{l}{\textsuperscript{*}\footnotesize Hardware variability refers to unit-to-unit performance variation among identical products.} \\
\end{tabularx}}
\end{table*}

\section{Background}\label{sec:background}

LLM serving exhibits variability from multiple sources. Table~\ref{tab:related-work} summarizes these including hardware variability from device-level performance differences due to process, power, and thermal effects, phase variability from differences between prefill and decode execution, model variability from input-dependent behavior such as MoE routing and expert activation skew, and service variability from request-level dynamics such as arrival rate, output length, and SLO heterogeneity. These factors interact to create execution-time imbalance across GPUs, directly affecting throughput and tail latency under synchronized execution. 

\begin{figure}
    \centering
    \begin{subfigure}[t]{0.359\linewidth}
        \centering
        \includegraphics[width=\linewidth]{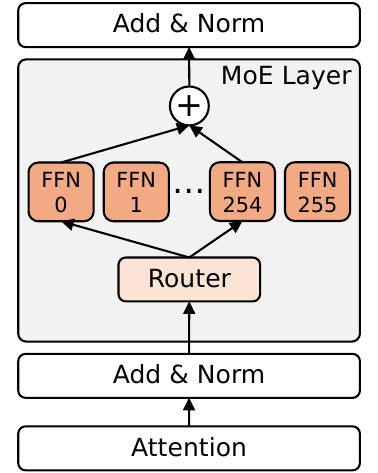}
        \caption{Single-GPU execution of an MoE model with 256 experts.}
        \label{fig:moe-single-gpu}
    \end{subfigure}
    \hfill
    \begin{subfigure}[t]{0.54\linewidth}
        \centering
        \includegraphics[width=\linewidth]{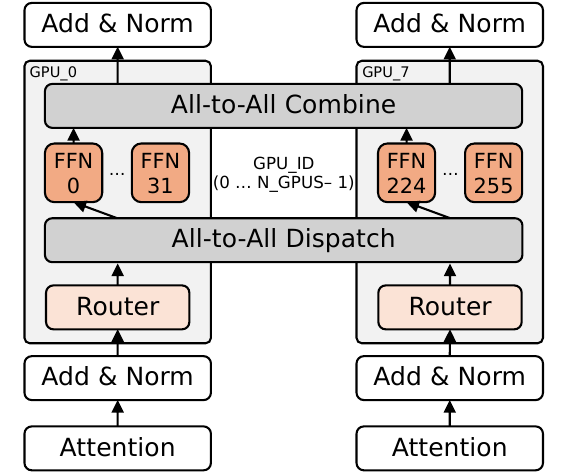}
        \caption{Expert-parallel execution across 8 GPUs, requiring all-to-all communication and synchronization.}
        \label{fig:moe-expert-parallel}
    \end{subfigure}
    \vspace{-0.5em}
    \caption{Comparison of MoE execution strategies: (a) single-GPU vs (b) expert-parallel execution.}
    \label{fig:moe_architecture}
\end{figure}

\niparagraph{Mixture-of-Experts and Expert Parallelism.}
Mixture-of-Experts (MoE) models enable conditional computation by routing each token to a subset of experts~\cite{deepspeed-moe,gshard,switchtransformer,deepseek,qwen,gptoss,shazeer2017outrageously}. This decouples total model capacity from per-token computation, allowing model size to scale without a proportional increase in compute. Instead of activating all parameters for every token, MoE layers activate only the selected experts, which makes sparse scaling practical for large language models. Figure~\ref{fig:moe-single-gpu} illustrates the conceptual single-GPU case, where all experts are placed on one device and token dispatch remains local. At the same time, routing is input-dependent, so expert activation can vary across requests and layers, creating uneven workloads across experts.

However, as MoE models grow, experts are distributed across GPUs using expert parallelism (EP), which reduces per-device memory footprint and allows the full expert set to fit across the system~\cite{li2023accelerating,fastermoe,fastmoe,moetuner,exflow,comet,megascaleinfer,megatron,vllm,deepspeed-moe}. 
As shown in Figure~\ref{fig:moe-expert-parallel}, EP partitions experts across GPUs, requiring token exchange through all-to-all communication before and after expert computation. 
This changes MoE execution from the local setting in Figure~\ref{fig:moe-single-gpu} to a distributed execution model in which layer latency is determined by the GPU that finishes last. As a result, performance becomes sensitive to both workload imbalance and device-level performance differences: routing skew can assign more work to some GPUs, while hardware variability can cause identical workloads to execute at different speeds. In both cases, faster GPUs idle while waiting at synchronization points.

\niparagraph{GPU Performance Variability.}
GPU performance variability can arise from several sources, including process variation, power-delivery constraints, temperature variations, and less common infrastructure-level effects~\cite{notallgpus,pal,litsilicon}. At the hardware level, nominally identical GPUs can exhibit different effective throughput because device-specific leakage and dynamic power characteristics determine the achievable operating frequency under a fixed power budget. These differences are exposed through dynamic voltage and frequency scaling (DVFS) and related control mechanisms. As workloads approach the power envelope, GPUs must satisfy board-level power budgets and per-rail current constraints, which can reduce attainable frequency and create performance asymmetry across devices.

Runtime conditions further amplify this. Higher temperatures increase leakage power, reducing the dynamic power headroom available for computation and increasing the probability of reaching the GPU power budget. Cooling asymmetry, rack-level temperature variation, and workload intensity can  change how strongly variability is expressed during execution. 
In MoE serving, the most relevant and common form of variability is performance divergence that appears when workloads push GPUs near their power or current limits.

\noindent \textbf{Prior Approaches to Mitigate Workload Skew and Hardware Variability.}
As summarized in Table~\ref{tab:related-work}, prior work addresses hardware, phase, model, and service variability in isolation and does not jointly optimize expert placement under both workload skew and hardware variability.
MoE-specific approaches such as EPLB~\cite{eplb} and ExFlow~\cite{exflow} optimize expert placement based on routing frequency or communication affinity, targeting token imbalance or communication cost. In contrast, hardware-aware approaches such as PAL~\cite{pal}, TAPAS~\cite{tapas}, Perseus~\cite{perseus}, and throttLL’eM~\cite{throttllem} address power or thermal asymmetry, typically through coarse-grained scheduling or frequency control. DynamoLLM~\cite{dynamollm} adapts to service-level variation by adjusting parallelism and operating frequency. 
These approaches improve individual dimensions of variability, but none jointly optimize expert placement for execution-time balance under both workload skew and heterogeneous GPU performance.

\section{Key Challenges}
\label{sec:challenges}

End-to-end performance of distributed MoE serving is determined by the interaction between \emph{workload skew} and \emph{hardware performance variability}. Existing load-balancing techniques usually address these factors separately, either by balancing token distribution or by improving device utilization, but not by jointly minimizing execution-time imbalance~\cite{tapas,sarathi,eplb}. In this section, we evaluate these factors running DeepSeek-V3 using 8$\times$EP on an 8-GPU \mithreetwentyfive system using the Sonnet dataset configured as described in the evaluation setup, Section~\ref{sec:eval}.

\begin{figure}[t]
\centering
\includegraphics[width=0.9\linewidth]{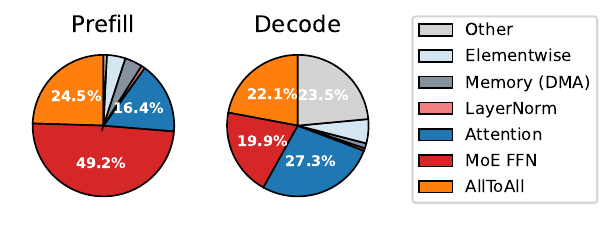}
\vspace{-1em}
\caption{Kernel latency breakdown for end-to-end inference run, using perfect token load balance across experts.}
\label{fig:time_breakdown}
\end{figure}

Our first observation is that MoE layers are the dominant target for optimization. As shown in Figure~\ref{fig:time_breakdown}, MoE FFN kernels account for 49\% of prefill and 20\% of decode time in DeepSeek-V3, while all-to-all contributes an additional 24.5\% and 22.1\%, respectively. Even under perfect token load balance, MoE layers dominate end-to-end latency.
Second, hardware variability manifests when the system is power constrained, and MoE layers drive GPU compute resources more intensively than attention, pushing devices closer to their power limits and thereby amplifying inter-GPU performance asymmetry. For example, for requests with 1,024-token inputs at 16 batch size, the MoE layer operates at the TDP limit for 82.8\% of its execution time, compared to only 34.8\% for the attention layer. This sustained power saturation reduces GPU frequency by 38\% on average for MoE layers, versus 10\% for attention, making MoE layers the dominant source of execution-time imbalance.
As such, MoE layers provide a direct lever for mitigation: routing and expert placement determine how work is distributed across GPUs. This makes it possible to offset hardware asymmetry through variability-aware placement. We therefore focus on MoE layers as both the dominant source of variability and the most effective target for execution-time balancing.

\begin{figure}[t]
    \centering
    \begin{subfigure}[t]{0.495\linewidth}
        \centering
        \includegraphics[width=\linewidth]{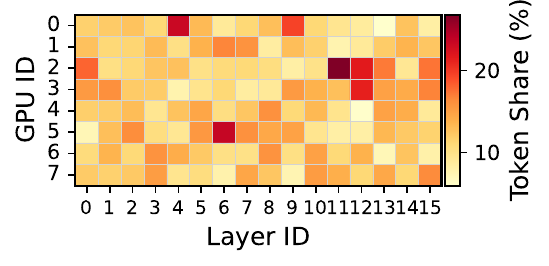}
        \vspace{-2em}
        \caption{Prefill token distribution.}
        \label{fig:expert_skew_prefill}
    \end{subfigure}
    \hfill
    \begin{subfigure}[t]{0.495\linewidth}
        \centering
        \includegraphics[width=\linewidth]{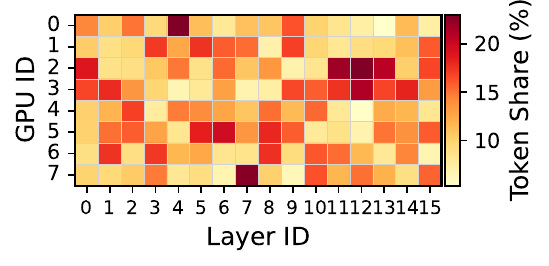}
        \vspace{-2em}
        \caption{Decode token distribution.}
        \label{fig:expert_skew_decode}
    \end{subfigure}
    \vspace{-0.5em}
    \caption{Token load distribution across GPUs for Sonnet on DeepSeek-V3 MoE using contiguous expert placement.
    }
    \label{fig:expert_skew}
\end{figure}

\niparagraph{Challenge 1: Routing skew creates shifting stragglers.}
MoEs route each token to a subset of experts, producing input-dependent compute and communication patterns. With expert parallelism, each GPU hosts the same number of experts, but the token load to those experts can vary substantially across layers. Figure~\ref{fig:expert_skew} shows the resulting per-GPU token distribution for DeepSeek-V3 with 256 experts under 8$\times$EP using vLLM contiguous expert placement. Although each GPU hosts 32 experts, token load is uneven; in layer 11 during prefill, the busiest GPU processes over 24\% of the tokens, while the least-loaded GPU handles less than 10\%.

Moreover, the degree of skew and the GPU with the most load change across layers. As a result, the bottleneck shifts during a single execution, with different GPUs becoming the straggler at different layers. Because all GPUs must synchronize at every MoE layer, this routing skew causes lightly loaded GPUs to idle while waiting for the heavily loaded GPUs to finish, reducing throughput and increasing tail latency.

\begin{tcolorbox}[
  colback=white, colframe=black,
  boxrule=0.5pt, arc=0pt,
  left=1pt, right=1pt, top=1pt, bottom=1pt
]
\textbf{\textit{Challenge: Token imbalance creates persistent stragglers.}}
Expert routing produces input-dependent load imbalance, leading to uneven work distribution across GPUs. Under expert parallelism, all GPUs must synchronize at each layer, so execution time is determined by the slowest device.
\\
\noindent \textbf{\textit{Insight: Expert placement provides a lever to reduce imbalance.}}
Expert placement can be used to redistribute workload and reduce per-layer imbalance without modifying model semantics.
\end{tcolorbox}

\niparagraph{Challenge 2: Hardware variability creates execution-time imbalance even with balanced token load.}
Even when token counts are balanced, nominally identical GPUs do not always execute the same workload at the same speed. As discussed in Section~\ref{sec:background}, this divergence arises from process variation across dies, which eventually manifests as frequency variations across GPUs~\cite{rahalarabi2024optimizing}.

\begin{figure}
    \centering
    \begin{subfigure}[t]{0.495\linewidth}
        \centering
        \includegraphics[width=\linewidth]{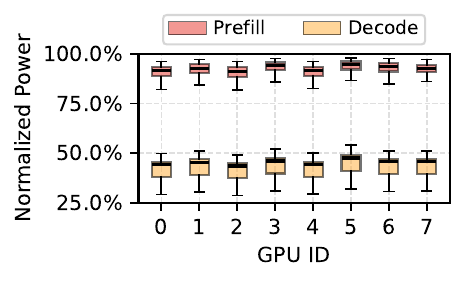}
        \vspace{-2em}
        \caption{Power distribution.}
        \label{fig:mi325x_power}
    \end{subfigure}
    \hfill
    \begin{subfigure}[t]{0.495\linewidth}
        \centering
        \includegraphics[width=\linewidth]{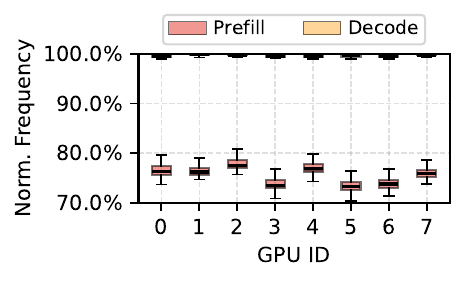}
        \vspace{-2em}
        \caption{Frequency distribution}
        \label{fig:mi325x_sclk}
    \end{subfigure}
    \vspace{-1em}
    \caption{GPU power and clock frequency distribution during DeepSeek-V3 prefill and decode. Power is normalized by TDP, and frequency is normalized by peak clock frequency.}
    \label{fig:telemetry_skew}
\end{figure}

Process variations matter when GPUs operate near their power limits. Under such conditions, DVFS enforces a fixed power budget, and device-specific power characteristics determine the achievable operating frequency. 
As a result, the same MoE kernel can execute at different speeds across GPUs.
Figure~\ref{fig:telemetry_skew} illustrates this mechanism. During prefill, MoE kernels have high compute requirements, power draw approaches TDP, and clock frequencies diverge across devices. During decode, power remains well below TDP, DVFS does not significantly constrain frequency, and clocks are correspondingly more uniform. This shows that hardware variability is a fixed property, but its performance impact is activated by workload intensity.

\begin{figure}
\centering
\includegraphics[width=0.9\linewidth]{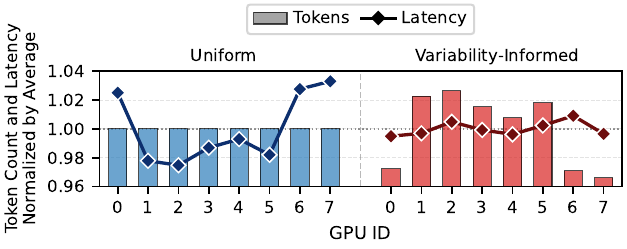}
\vspace{-1em}
\caption{Token load and MoE kernel latency per GPU under uniform and variability-informed token assignment. All values are normalized by the cross-GPU average.}
\label{fig:eplb_valb_kerneltime}
\end{figure}

To isolate the consequence of this effect, Figure~\ref{fig:eplb_valb_kerneltime} compares two controlled token assignments for a single MoE layer, with data normalized to the average across 8 GPUs. In the first chart, tokens are distributed uniformly across GPUs. In the second, knowing the measured per-GPU performance characteristics, we assign proportionally more tokens to faster GPUs and fewer tokens to slower GPUs.
Even in this straightforward variability-informed assignment we observe better balance across per-GPU completion times.

\begin{tcolorbox}[
  colback=white, colframe=black,
  boxrule=0.5pt, arc=0pt,
  left=1pt, right=1pt, top=1pt, bottom=1pt
]
\textit{\textbf{Challenge: GPU performance diverges under power-limited execution.}}
Nominally identical GPUs do not sustain the same operating frequency when constrained by a similar power budget, leading to execution-time differences across devices even under balanced token load.
\\
\noindent \textit{\textbf{Insight: Measured per-GPU performance must guide placement.}}
Device-specific performance characteristics can be profiled and incorporated into placement decisions to better match workload to effective compute capability.
\end{tcolorbox}

\niparagraph{Challenge 3: Placement becomes stale under workload/performance drift.}
The interaction between routing skew and hardware variability is not fixed over time. In practice, both workload intensity and effective GPU performance can drift across serving conditions. Changes in batch token count, input mix, or execution phase alter the stress level of the MoE kernels, which in turn changes the degree of hardware-induced performance asymmetry. As a result, a placement that is well-matched to one serving scenario can become suboptimal as serving conditions evolve.

Prior work such as EPLB provides basic dynamic recalibration, but it remains limited in both \emph{how} and \emph{when} it adapts. First, recalibration is driven by token distribution ratios across experts or GPUs without considering the absolute token magnitude. This is insufficient because hardware variability is stress-dependent: two batches may exhibit the same routing ratio while exposing different levels of power-constrained latency variability. Second, recalibration is typically performed at fixed intervals, such as every several hundred or thousand iterations, regardless of whether the workload has actually changed. This incurs reshuffling overhead even when the existing placement remains appropriate while failing to respond promptly when drift is significant.

Our observation is that recalibration should be both \emph{stress-aware} and \emph{workload-aware}. To determine \emph{how} to recalibrate, the system must account for token magnitude since it directly affects the level of workload-induced hardware variability and therefore the performance asymmetry across GPUs. To determine \emph{when} to recalibrate, the system should monitor drift in batch token count and observed latency imbalance, and refresh the placement only when that drift exceeds a configurable threshold. This avoids unnecessary reshuffling while preserving alignment between workload distribution and effective device performance.

\begin{tcolorbox}[
  colback=white, colframe=black,
  boxrule=0.5pt, arc=0pt,
  left=1pt, right=1pt, top=1pt, bottom=1pt
]
\textit{\textbf{Challenge: Static or periodic placement updates become stale under workload/performance drift.}}
Token routing ratios alone do not capture the workload stress that triggers hardware variability, and fixed-interval recalibration can incur reshuffle overhead while missing meaningful drift.
\\
\noindent\textit{\textbf{Insight: Recalibration should be drift-triggered.}}
Placement should be refreshed using both token numbers and latency imbalance when workload/performance drift exceeds a configurable threshold.
\end{tcolorbox}

\begin{figure*}
\begin{center}
\includegraphics[width=1\textwidth]{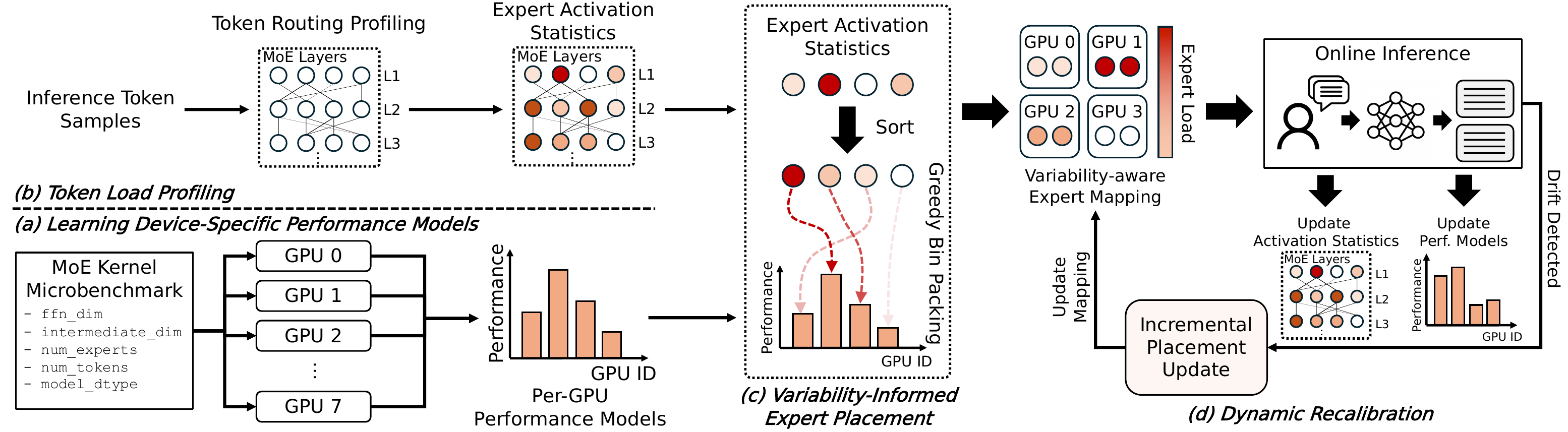}
\end{center}
\caption{\seokjin{\ours framework: (a) learning device-specific performance models, (b) characterization of expert activation patterns, (c) variability-informed expert placement that minimizes per-layer execution-time imbalance, and (d) dynamic recalibration of expert mapping when workload drift is detected.}}
\label{fig:ViBE}
\end{figure*}

\section{\ours Framework}

We present \ours, a variability-aware expert placement framework that minimizes execution-time imbalance in MoE inference by jointly reasoning about workload skew and hardware performance.

The core challenge is that execution time depends on two factors that evolve differently over time: input-dependent expert load and device-specific performance. 
The system assigns experts to GPUs, but latency emerges from a complex interaction between token load, device-specific performance, and synchronization, requiring accurate prediction rather than direct optimization.
Moreover, the underlying workload and effective GPU throughput can drift, making a one-time placement suboptimal. 

\ours addresses this challenge through four components as shown in Figure~\ref{fig:ViBE}. %
First, it builds device-specific execution models that map token load to latency. 
Second, it profiles expert activation patterns to estimate workload skew for each layer. 
Third, it computes an execution-time-aware placement that aligns predicted per-layer completion times across GPUs. 
Fourth, it supports drift-aware incremental updates, modifying placement only when workload or performance characteristics have changed enough to affect balance.
By integrating these components, \ours shifts the objective from token balancing to execution-time balancing, improving predictability and enabling higher system utilization.
The end-to-end flow of \ours is formalized in Algorithm~\ref{alg:vibe}; the remainder of this section details each component.

\subsection{Design Principles}

\niparagraph{Balance execution time, not nominal load.}
Hardware variability can arise from multiple sources, including process variation, power-delivery constraints, and temperature. Regardless of the source, its system-level consequence is the same: GPUs assigned similar work can still complete at different times, creating latency imbalance under synchronized expert-parallel execution. In this work, we focus on the primary and more common case in MoE serving, where GPUs are pushed toward their power limits and performance asymmetry is exposed through DVFS. While this power-limited regime is the main source of variability we study, the need to balance execution time rather than nominal load applies broadly to other forms of hardware-induced variability.

\textit{Design Principle 1: Optimize for aligned completion time.
Because hardware variability causes some GPUs to finish later than others, expert placement should not optimize for equal token counts alone, but for aligned execution time across GPUs.}

\niparagraph{Reduce execution-time outliers.}
With expert parallelism, each MoE layer executes in a synchronized manner across GPUs, thus performance is governed by the tail of the execution time distribution rather than by average load. As a result, small differences in workload or device performance create stragglers that dominate layer latency and accumulate across execution. Figure~\ref{fig:eplb_valb_kerneltime} illustrates that uniform token assignment still produces a spread in kernel execution times, whereas variability-aware assignment reduces this spread by aligning per-GPU completion time.

\textit{Design Principle 2: Minimize the slowest-GPU penalty.
We reveal a mismatch in token-balanced-only objectives: equalizing token counts does not eliminate stragglers, hence, the placement objective should therefore minimize execution-time outliers, not just token-count imbalance.}

\subsection{Framework Design}

\subsubsection{Device-Specific Performance Models}
Nominally identical GPUs exhibit differences in effective kernel throughput due to process variation and power-limited operation. Under expert parallelism (EP), these differences directly translate into execution-time imbalance, as each layer is gated by the slowest device.

To enable latency-aware placement, we construct per-GPU performance models that capture the relationship between compute load and latency. 
Each GPU is warmed to steady-state operating conditions and profiled independently using the fused MoE kernel across a range of token counts. This produces a function $f_g(n)$ that maps token load $n$ to expected kernel latency. 
The relationship between token load and expected latency per GPU does not vary over time~\cite{pal}; hence profiling can be done once per GPU and retained during normal operation.

\subsubsection{Token Load Profiling.}
MoE routing produces input-dependent and layer-specific workload distributions, making it difficult to determine expert load a priori. However, empirical observations show that activation patterns are relatively stable for a given benchmark.
We leverage this property by profiling expert activation over a representative input set. This produces an activation matrix $\mathbf{W} \in \mathbb{R}^{L \times E}$, where each entry $w_e^{(l)}$ captures the relative load of expert $e$ at layer $l$.

\subsubsection{Variability-Informed Expert Placement}
We formalize expert placement as an execution-time minimization problem.
The central idea is to assign experts so that projected execution times---not token
counts---are equalized across GPUs within each EP group.

\niparagraph{Problem Formulation.}
Each MoE layer assigns $E$ logical experts to $G$ GPUs.
Given a placement $\{S_g\}$, the token load on GPU~$g$ is $n_g = \sum_{e \in S_g} w_e$,
and the layer latency is determined by the slowest device:
\[
    T = \max_{g} \; f_g(n_g),
\]
and the objective minimizes $\sum_l T^{(l)}$ across all $L$ MoE layers.

\niparagraph{Latency-aware solver.}
Because the optimal partition problem is NP-hard in general (unrelated-machines makespan minimization~\cite{lenstra1990approximation}), \ours applies a per-layer greedy heuristic. For each GPU, we derive a speed estimate $s_g = 1 / f_g(n_{\mathrm{ref}})$ from the performance model, where $n_{\mathrm{ref}}$ is the mean per-expert token load, then define a token target
\[
    \tau_g = N \cdot \frac{s_g}{\sum_{h} s_h},
\]
where $N = \sum_e w_e$ is the layer's total token load, encoding how much load each GPU should receive so that all devices finish at the same predicted latency. 
Experts are assigned in descending order of $w_e$ to the GPU farthest below its target~$\tau_g$.
This enables faster GPUs to absorb high-load experts, directly equalizing predicted execution time rather than token count.
In contrast, EPLB implicitly assumes $f_g(n) = n$, so it cannot compensate for hardware throughput differences.

\subsubsection{Dynamic Recalibration}

\seokjin{
Effective throughput can shift over time due to changes in workload characteristics such as batch size, token distribution, and execution phase (e.g., prefill versus decode). To enable the system to adapt to changing operating conditions, \ours supports lightweight recalibration of the observed per-expert token load $n$ fed into the performance model $f_g$. 
}

\seokjin{
Prior work rearranges experts every $k$ steps by re-running a full greedy solver~\cite{eplb}, which has two limitations.
First, it balances token \emph{ratios} but ignores \emph{magnitude}: a request-rate spike shifts GPUs into steeper regions of $f_g(n)$, changing the optimal placement even when token ratios are unchanged.
Second, fixed-cadence rearrangement wastes resources during steady state, reacts late during transients, and shuffles most experts even when a small adjustment suffices.
\ours addresses both gaps.
}

\niparagraph{Drift-triggered recalibration.}
Because \ours tailors expert placement to the observed routing pattern --- assigning high-load experts to faster GPUs --- the placement becomes suboptimal when the input distribution shifts and different experts become highly loaded.
To detect such shifts, \ours tracks the cosine similarity between the current and reference per-layer expert load vectors.
After each rearrangement, it records the mean load vector $\hat{\mathbf{w}}$, averaged over a window of 100 samples at every MoE layer. 
Every $H$ forward passes (default $H{=}10$), \ours computes the cosine distance between the current 100-sample window average $\mathbf{w}$ and the reference:
\[
    d \;=\; 1 - \frac{\mathbf{w} \cdot \hat{\mathbf{w}}}
                      {\|\mathbf{w}\|\;\|\hat{\mathbf{w}}\|}.
\]
When any layer's $d$ exceeds a threshold $\delta_{\cos}$ (default $0.05$), the routing pattern has diverged enough to warrant readjustment, and \ours triggers the incremental solver (Algorithm~\ref{alg:incremental}) with minimal expert movement.

\niparagraph{Incremental placement update.}
\seokjin{
Rather than re-solving from scratch, \ours starts from the current placement and applies the minimum number of cross-GPU expert swaps to restore balance (Algorithm~\ref{alg:incremental}).
At each iteration, the solver identifies the GPU pair with the largest latency gap and evaluates all valid swaps between them, scoring each by its marginal latency gain using the device-specific $f_g$.
The process terminates once all GPUs fall within tolerance $\epsilon$ (default 0.03) of the target latency:
\[
    \max_g f_g\!\left(n_g\right)
    \;\leq\;
    (1 + \epsilon)\;\cdot\;
    \frac{1}{|G|}\sum_g f_g\!\left(n_g\right).
\]
In practice, full rebalancing by EPLB or \ours's initial solver reassigns over 200 of the 256 physical expert slots per layer, incurring substantial weight-transfer overhead. The incremental solver typically converges in 5--30 swaps per layer, reducing transfer volume by over an order of magnitude.
}

\section{Evaluation}
\label{sec:eval}

\subsection{Experimental Setup}

\niparagraph{Hardware and Software Platform.}
We evaluate ViBE using vLLM~v0.14.2~\cite{vllm}, PyTorch v2.9.0~\cite{pytorch} on ROCm~v7.0~\cite{rocm}.
We use AITER~\cite{aiter} as the attention and MoE kernel backend.
Experiments are conducted on a single-node platform with 8$\times$ \mithreetwentyfive GPUs,
each featuring up to 2100\,MHz clock frequency, 256\,GB HBM3e memory, and a 1000\,W TDP~\cite{mi325x}.
\seokjin{
All measurements are collected on single-node deployments to isolate intra-node variability effects.
We use rocprofv3~\cite{rocprof} and PyTorch Profiler~\cite{torchprof} to profile kernel traces, and ROCm SMI~\cite{rocmsmi} to measure hardware telemetry.
}

\begin{table}[t]
\centering
\caption{Evaluation Setup.}
\label{tab:eval-setup}
\begin{subtable}[t]{\columnwidth}
\centering
\caption{Models and Configurations.}
\label{tab:eval-models}
\resizebox{0.85\columnwidth}{!}{
\begin{tabular}{|l|c|c|c|c|}
    \hline
    \textbf{Model} & \textbf{dtype} & \textbf{\# Experts} & \textbf{EP Degree} & \textbf{Routed Experts} \\
    \hline
    DeepSeek-V3~\cite{deepseek} & FP8 & 256 & 8 & 8 \\
    \hline
    Qwen-3 235B~\cite{qwen} & FP8 & 128 & 8 & 8 \\
    \hline
\end{tabular}}
\end{subtable}

\begin{subtable}[t]{\columnwidth}
\centering
\caption{SLO Thresholds by Benchmark and Model.}
\label{tab:slo-targets}
\resizebox{0.95\columnwidth}{!}{
\begin{tabular}{|l|c|c|c|l|c|c|}
    \hline
    \textbf{Dataset} & \textbf{Input Len.} & \textbf{Output Len.} & \textbf{\# Req.} & \textbf{Model} & \textbf{TTFT SLO} & \textbf{TPOT SLO} \\
    \hline
    \multirow{2}{*}{ShareGPT} & \multirow{2}{*}{219.2 (Avg.)} & \multirow{2}{*}{200.8 (Avg.)} & \multirow{2}{*}{5K}
        & DeepSeek-V3   & 250ms & 125ms \\ \cline{5-7}
    & & & & Qwen-3 235B & 250ms & 100ms \\ \hline
    \multirow{2}{*}{Sonnet} & \multirow{2}{*}{1024 (Fixed)} & \multirow{2}{*}{128 (Fixed)} & \multirow{2}{*}{1K}
        & DeepSeek-V3   & 350ms & 125ms \\ \cline{5-7}
    & & & & Qwen-3 235B & 350ms & 100ms \\ \hline
\end{tabular}}
\end{subtable}

\begin{subtable}[t]{\columnwidth}
\centering
\caption{Expert Placement Policies.}
\label{tab:placement-policies}
\resizebox{0.95\columnwidth}{!}{
\begin{tabular}{|l|l|c|}
    \hline
    \textbf{Policy} & \textbf{Objective} & \textbf{Hardware Aware?} \\
    \hline
    vLLM~\cite{vllm,deepspeed-moe}
        & Contiguous expert partitioning
        & No \\
    \hline
    EPLB~\cite{deepseek,eplb}
        & Equalize token load across GPUs
        & No \\
    \hline
    ViBE
        & Minimize execution time imbalance
        & Yes \\
    \hline
\end{tabular}}
\end{subtable}

\end{table}

\niparagraph{Models and Parallelism Configuration.}
All experiments use hybrid Tensor Parallelism (TP) and Expert Parallelism (EP) across 8 GPUs. For dense layers, we use TP degree 8 and for each MoE layer we use EP degree 8, shown in Table~\ref{tab:eval-models}. 
Thus, for MoE layers, all GPUs participate in lockstep to process experts, whereas non-MoE layers use standard tensor parallelism. 
\seokjin{
Although \ours is compatible with alternative parallelism choices, including DP for non-MoE layers, we use EP+TP to isolate variability from expert load imbalance. \ours remains applicable to EP+DP as well.
}

\niparagraph{Workloads and Metrics.}
We evaluate two representative benchmarks~\cite{dynaserve,distserve}, listed in Table~\ref{tab:slo-targets}. Requests are replayed at fixed target rates with a Poisson arrival process using the vLLM client.
Modern LLM serving systems often use prefill--decode disaggregation~\cite{distserve,splitwise}. To emulate this scenario, we evaluate prefill and decode separately using our experimental node. Prefill is isolated with long-input, single-output-token requests (e.g., 1024-in / 1-out for Sonnet), while decode is isolated by warming the prefix cache so that measurement skips prefill computation. During decode measurements, prefix-cache hit rate remained at 100\% except for the final 16-token block, which vLLM recomputes to produce logits for the first decode step.\footnote{Under disaggregation, KV cache transfer is orthogonal to expert placement and identical across all evaluated strategies.}
For each model and hardware platform, we sweep target request rates (QPS) and report TTFT and TPOT percentiles, capturing both interactive latency and sustained generation performance under load.

\niparagraph{Service-Level Objective (SLO).}
We define per-model and per-platform SLO thresholds reflecting realistic interactive-serving requirements (Table~\ref{tab:slo-targets}), accounting for differences in model size and datasets while being consistent with prior serving systems~\cite{dynamollm,dynaserve}.
We report \emph{goodput}---the rate of SLO-compliant requests~\cite{distserve}---as the primary quality-of-service metric.  
We target 90\% goodput and report the maximum sustainable QPS that maintains this compliance level.

\niparagraph{Expert Placement Policies.}
We compare three expert placement strategies that differ in their optimization objective and hardware awareness (Table~\ref{tab:placement-policies}).
All placements are computed offline from profiling data and held static during serving to isolate the impact of the placement policy itself. \seokjin{In all three configurations, we assign the same number of experts per GPU to ensure uniform memory usage. Note, however, that this is not strictly required by either EPLB or \ours. We leave non-uniform expert allocation to future work. }

\subsection{Overall Results}\label{sec:results}

\begin{figure}[t]
    \centering
    \begin{subfigure}[b]{0.45\linewidth}
        \centering
        \includegraphics[width=\linewidth]{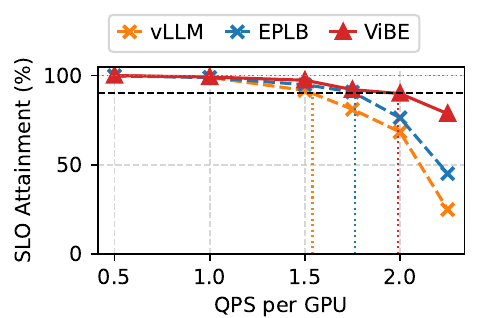}
        \caption{\seokjin{DeepSeek-V3 - Sonnet.}}
        \label{fig:mi325x-deepseek-slo}
    \end{subfigure}
    \hfill
    \begin{subfigure}[b]{0.45\linewidth}
        \centering
        \includegraphics[width=\linewidth]{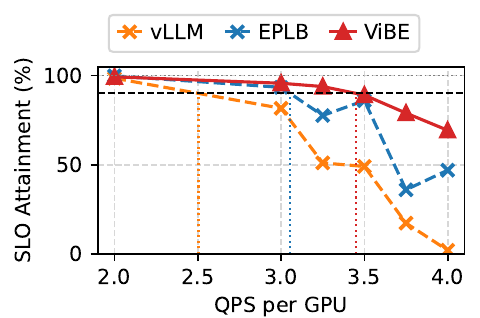}
        \caption{\seokjin{Qwen-3 - Sonnet.}}
        \label{fig:mi325x-qwen-slo}
    \end{subfigure}

    \begin{subfigure}[b]{0.45\linewidth}
        \centering
        \includegraphics[width=\linewidth]{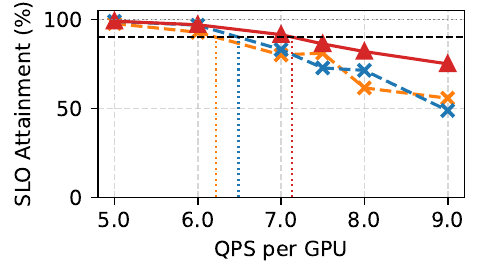}
        \caption{\seokjin{DeepSeek-V3 - ShareGPT.}}
        \label{fig:mi300x-deepseek-slo}
    \end{subfigure}
    \hfill
    \begin{subfigure}[b]{0.45\linewidth}
        \centering
        \includegraphics[width=\linewidth]{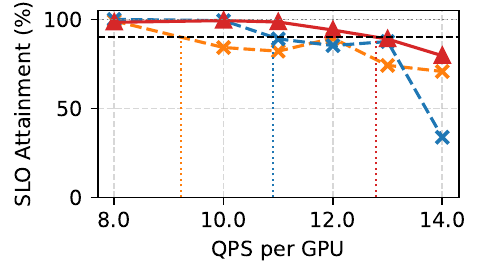}
        \caption{\seokjin{Qwen-3 - ShareGPT.}}
        \label{fig:mi300x-qwen-slo}
    \end{subfigure}

    \caption{SLO attainment across request rates.}
    \label{fig:slo-attainment}
\end{figure}

\begin{figure*}[t]
    \centering

    \begin{minipage}{1\textwidth}
        \centering
        \begin{subfigure}[b]{0.48\linewidth}
            \centering
            \includegraphics[width=\linewidth]{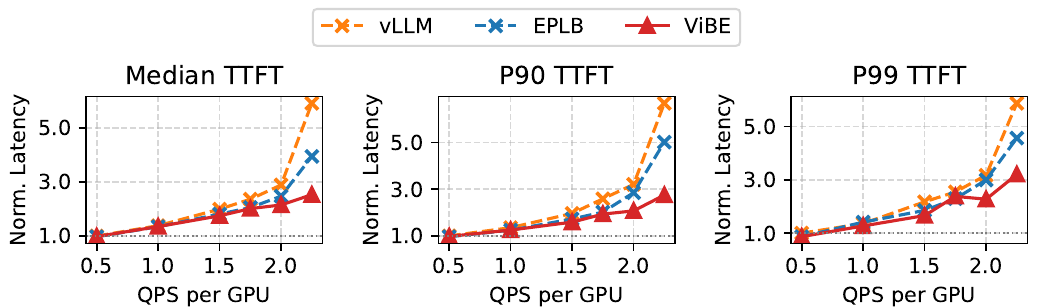}
            \includegraphics[width=\linewidth]{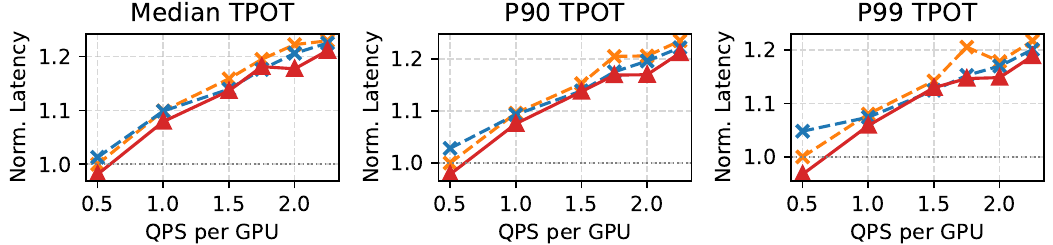}
            \caption{\seokjin{Key performance results of DeepSeek-V3 for Sonnet dataset.}}
            \label{fig:mi325x-deepseek}
        \end{subfigure}
        \hfill
        \begin{subfigure}[b]{0.48\linewidth}
            \centering
            \includegraphics[width=\linewidth]{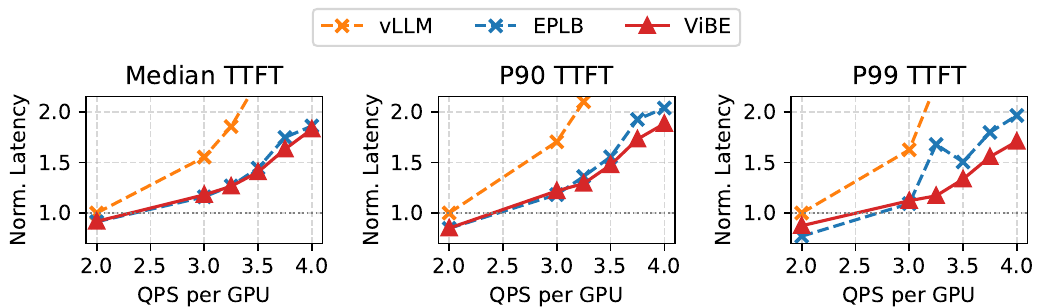}
            \includegraphics[width=\linewidth]{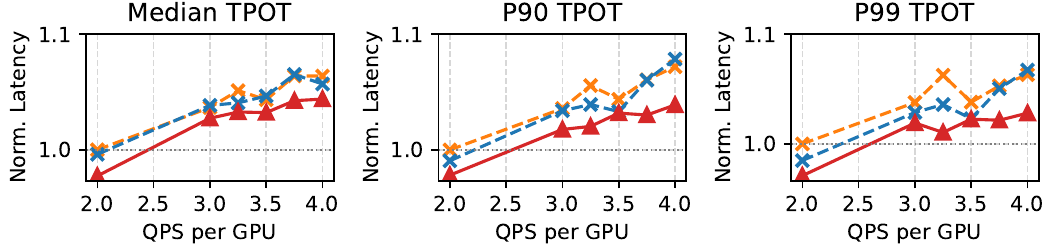}
            \caption{\seokjin{Key performance results of Qwen-3 for Sonnet dataset.}}
            \label{fig:mi325x-qwen}
        \end{subfigure}
    \end{minipage}

    \begin{minipage}{1\textwidth}
        \centering
        \begin{subfigure}[b]{0.48\linewidth}
            \centering
            \includegraphics[width=\linewidth]{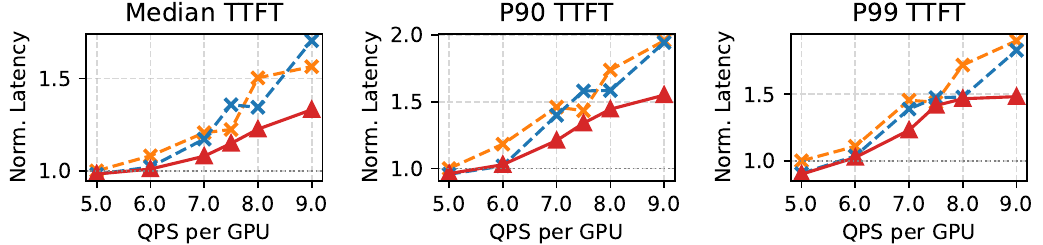}
            \includegraphics[width=\linewidth]{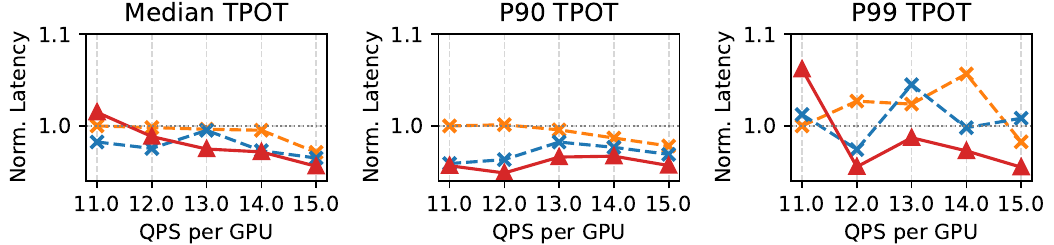}
            \caption{\seokjin{Key performance results of DeepSeek-V3 for ShareGPT dataset.}}
            \label{fig:mi325x-deepseek-sharegpt}
        \end{subfigure}
        \hfill
        \begin{subfigure}[b]{0.48\linewidth}
            \centering
            \includegraphics[width=\linewidth]{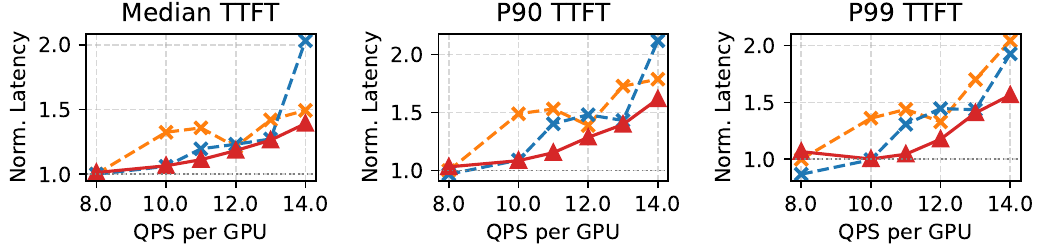}
            \includegraphics[width=\linewidth]{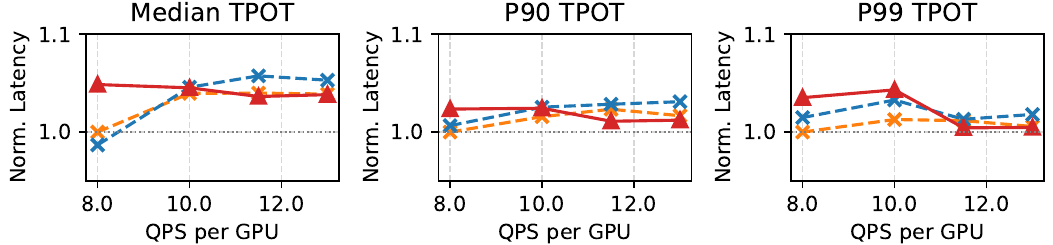}
            \caption{\seokjin{Key performance results of Qwen-3 for ShareGPT dataset.}}
            \label{fig:mi325x-qwen-sharegpt}
        \end{subfigure}
    \end{minipage}
    \caption{\seokjin{End-to-end performance of vLLM, EPLB, and ~\ours. All values are normalized by baseline vLLM at the lowest QPS.}}
    \label{fig:main-results}
\end{figure*}

\niparagraph{SLO Attainment.}
Figure~\ref{fig:slo-attainment} shows SLO attainment as a function of request rate across both models and datasets.

\seokjin{
On Sonnet, which uses fixed input lengths, goodput follows a consistent ordering---vLLM $<$ EPLB $<$ \ours---across all QPS (Figures~\ref{fig:mi325x-deepseek-slo},~\ref{fig:mi325x-qwen-slo}).
Fixed-length inputs produce stable routing patterns that closely match the time-averaged statistics used for placement, allowing each successive policy to deliver its expected benefit.
ShareGPT's variable-length inputs introduce more routing variance (Figures~\ref{fig:mi300x-deepseek-slo},~\ref{fig:mi300x-qwen-slo}): depending on batch scheduling and timing, hot experts can exhibit sudden load spikes that deviate from the profiled average.
EPLB, which balances only token ratios without hardware awareness, may assign these spike-prone experts to slow GPUs, narrowing or erasing its advantage over vLLM at certain QPS ranges.
\ours remains consistently ahead because it steers hot experts toward the fastest GPUs, absorbing routing variance through hardware headroom rather than relying solely on token balance.
This translates directly to reduced synchronization overhead: \ours lowers per-layer barrier idle time by 41\% over EPLB, yielding a 35\% reduction in total synchronized MoE execution latency.
Comparing across models, Qwen-3 is lighter than DeepSeek-V3 and sustains higher request rates before reaching the SLO limit.
This pushes the workload deeper into the compute-bound regime at high QPS, where per-GPU throughput differences are most pronounced---consistent with the variability characterization in Section~\ref{sec:challenges}.
As a result, the gap between placement policies is larger for Qwen-3: \ours extends the SLO frontier by \seokjin{15}\% relative to EPLB, compared to \seokjin{12}\% for DeepSeek-V3.
}

\niparagraph{Impact of Hardware Stress on Placement Effectiveness.} Figure~\ref{fig:main-results} shows the median, P90 and P99 TTFT and TPOT latency for the models and datasets. All results are normalized to the left-most QPS per GPU rate shown for vLLM. The impact of variability-aware placement scales with the operating regime of the system, whether that be because of model size, model phase (TTFT versus TPOT), or QPS rate. DeepSeek-V3, with larger expert dimensions, saturates GPU compute earlier, limiting the range of request rates feasible within the given SLO. 
Sweeping across QPS rates shows that when GPUs are underutilized, hardware variability is minimal and EPLB can match or slightly outperform \ours (e.g., Figure~\ref{fig:mi325x-qwen} at QPS~2.0). 
This is especially true for TPOT, where a combination of the lower intensity decode (Figure~\ref{fig:telemetry_skew}) combined with a light QPS load can result in higher TPOT for \ours. In this scenario, other types of resource contention between hot experts assigned to the same GPU (e.g., memory BW, communication BW) outweigh the benefits of \ours. 
As QPS increases and GPUs saturate, execution-time variability emerges, and \ours consistently outperforms both vLLM and EPLB.
This range is the most relevant in practice, as it determines the sustainable throughput that meets SLOs.
\ours mitigates stragglers, allowing the system to sustain higher request rates without violating latency targets.

\niparagraph{Tail Latency and User Interactivity.}
\seokjin{
The gains are most visible in the P90--P99 range: for DeepSeek-V3 on Sonnet dataset, \ours reduces P90 and P99 TTFT by up to \seokjin{45\% and 30\%} respectively; for Qwen-3, by up to \seokjin{10\% and 30\%}. 
The larger gains on DeepSeek-V3 reflect its higher per-token compute, which amplifies straggler effects as QPS increases. These tail-latency reductions come without sacrificing throughput, showing that variability-aware placement addresses a bottleneck beyond token-count balancing.
}

\subsection{Kernel Time and Hardware Telemetry}
\seokjin{
To understand the mechanisms behind \ours's improvements in SLO and tail latency, we examine per-GPU execution characteristics at the MoE layer level, combining kernel timing breakdowns with hardware telemetry.
}

\begin{figure}
\centering
     \begin{subfigure}[b]{0.303\linewidth}
         \centering
         \includegraphics[width=\linewidth]{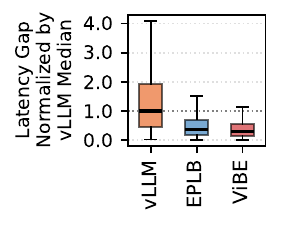}
         \caption{Latency gap.}
         \label{fig:moe-latency-gap}
     \end{subfigure}
     \hfill
     \begin{subfigure}[b]{0.687\linewidth}
         \centering
         \includegraphics[width=\linewidth]{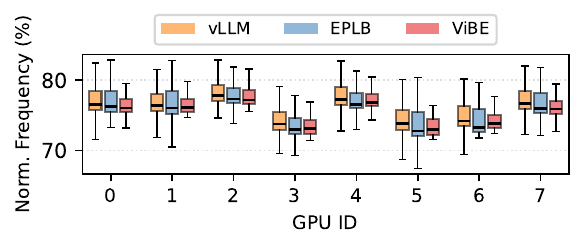}
         \caption{Normalized GPU clock frequency.}
         \label{fig:gpu-clock-balance}
     \end{subfigure}
    \caption{Performance variability during DeepSeek-V3 prefill: (a)~distribution of per-layer MoE kernel latency gap and (b)~clock frequency per GPU, normalized by peak.}
     \label{fig:hw-variability}
\end{figure}

\niparagraph{Kernel time variability.}
We measure the latency gap between the fastest and slowest GPU across each invocation of the MoE kernel when running DeepSeek-V3 prefill for Sonnet dataset at QPS 2.0. In the ideal case with no token or hardware variability, the latency gap should be zero meaning all GPUs complete at the same time.  Figure~\ref{fig:moe-latency-gap} shows the box plot of the latency gap for the three configurations: vLLM, EPLB, and \ours. Token redistribution reduces the median latency gap by 63.9\% for EPLB and an additional 19.6\% reduction is achieved with \ours. Balancing the work across GPUs and reducing the latency gap results in a 49.3\% and 27.9\% improvement in average MoE latency for \ours when compared to vLLM and EPLB, respectively.

\niparagraph{Hardware telemetry confirms even utilization.}
Figure ~\ref{fig:gpu-clock-balance} shows per-GPU clock frequency distribution during the same prefill workload.
Under vLLM and EPLB, GPUs exhibit a wide inter-device spread in mean frequency, reflecting uneven load across GPUs that does not take GPU hardware variability into consideration. 
\ours narrows this inter-GPU frequency gap, and also compresses the intra-GPU distribution: both the upper and lower tails of the per-GPU frequency box tighten, indicating that each device operates more consistently over time rather than alternating between bursts and idle periods.
This is consistent with the reduced latency gap -- more uniform frequency across GPUs reflects more uniform utilization across time.
Together, these two figures show that ~\ours's placement decisions reduce kernel time variance, suppress straggler idle time, and produce more stable per-GPU hardware utilization -- all of which compound into the end-to-end TTFT and TPOT improvements.

\subsection{Dynamic Workload Adaptation}\label{sec:dynamic}
\seokjin{
In production, traffic patterns change over time, creating  drift in token distribution and workload intensity, both leading to shifts in hardware variability behavior.
We evaluate how different recalibration strategies handle such drift using a workload transition benchmark.
}

\niparagraph{Setup.}
\seokjin{
To evaluate such workload shift, we derive expert placement from one dataset and deploy it on another. We consider ShareGPT$\to$Sonnet (SG$\to$SN) and Sonnet$\to$ShareGPT (SN$\to$SG), and compare them with the matched-workload cases SG$\to$SG and SN$\to$SN from Section~\ref{sec:results}. Static approach of EPLB and \ours retain the placement from original profiling, while adaptive updates the expert-to-GPU mapping online using measurements from the serving workload.
}

\begin{figure}[t]
  \centering
  \begin{subfigure}[t]{0.45\linewidth}
    \centering
    \includegraphics[width=\linewidth]{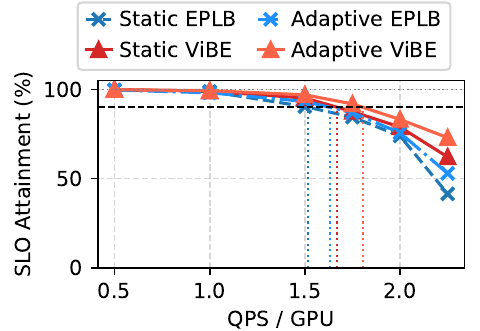}
    \caption{SG\(\to\)SN}
    \label{fig:xwork-slo-sg-sn}
  \end{subfigure}
  \hfill
  \begin{subfigure}[t]{0.45\linewidth}
    \centering
    \includegraphics[width=\linewidth]{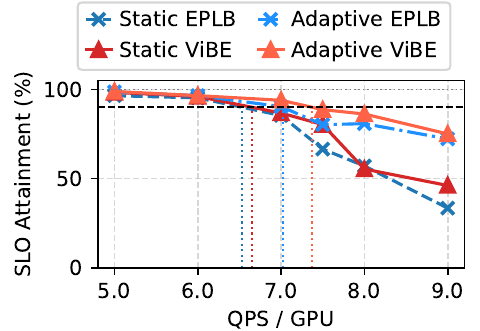}
    \caption{SN\(\to\)SG}
    \label{fig:xwork-slo-sn-sg}
  \end{subfigure}
  \caption{
SLO attainment across target QPS under cross-workload scenarios. Static placements degrade relative to matched-workload baselines because expert costs profiled on one workload do not transfer to another.
  }
  \label{fig:xwork-slo}
\end{figure}

\niparagraph{Serving quality.}
\seokjin{
Figure~\ref{fig:xwork-slo} shows TTFT SLO attainment under cross-workload drift.
Both static strategies see meaningful drops compared to the matched-workload baselines (SN$\to$SN, SG$\to$SG), which is expected: the expert execution profile captured during the profiling phase no longer reflects the actual serving workload, so the placement is suboptimal.
The adaptive variants substantially recover this gap.
In the SG$\to$SN scenario, Static \ours meets the 90\% SLO target up to 1.68 QPS/GPU, while Adaptive \ours extends this to 1.80 QPS/GPU; a similar pattern holds for EPLB (1.51 $\to$ 1.63 QPS/GPU).
The effect is even more pronounced in the SN$\to$SG direction: ShareGPT's variable input lengths and more diverse expert routing patterns create a larger distribution shift from the Sonnet profile, making recalibration more effective. Adaptive \ours pushes the 90\% crossover from 6.7 to 7.4 QPS/GPU, and Adaptive EPLB from 6.5 to 7.0 QPS/GPU.
Across both transition directions, the adaptive strategies bring SLO attainment closer to the levels achieved when the profiling and serving workloads match, demonstrating that periodic recalibration effectively tracks workload drift.
}

\begin{figure}
  \centering
  \includegraphics[width=0.98\linewidth]{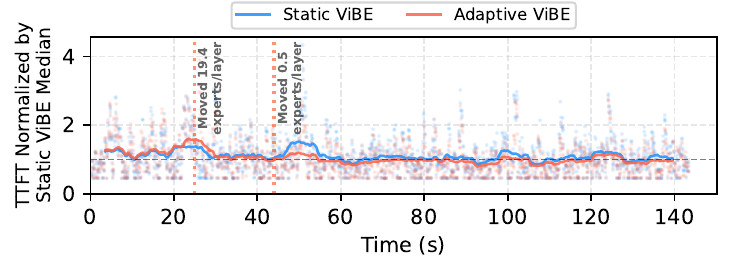}
  \caption{
    Per-request TTFT timeseries during the SG\(\to\)SN serving phase under adaptive recalibration. The lines represent rolling average with a window of 100 requests.
    Vertical dashed lines represent expert rearrangement events.}
  \label{fig:xwork-ttft-timeseries}
\end{figure}

\niparagraph{Recalibration overhead.}
\seokjin{
Adaptive recalibration is not free.
Figure~\ref{fig:xwork-ttft-timeseries} shows the per-request TTFT timeseries during adaptive serving.
Each recalibration event triggers an expert rearrangement phase in which a subset of experts is migrated between GPUs to match the updated placement.
During this migration window, requests that overlap with the rearrangement observe elevated TTFTs, visible as transient spikes in the timeseries.
These spikes are short-lived -- serving quality recovers within seconds once the new placement is installed -- but they do contribute to the remaining SLO attainment gap between adaptive cross-workload and matched-workload baselines.
The net effect is a trade-off: adaptive recalibration recovers the bulk of the performance lost to workload drift, at the cost of brief latency disruptions during each rearrangement event.
Reducing this migration overhead -- e.g., through incremental or speculative placement updates -- is an avenue for future work.
}

\subsection{Discussion}

\begin{figure}[t]

    \begin{subfigure}[t]{0.46\linewidth}
        \centering
        \includegraphics[width=\linewidth]{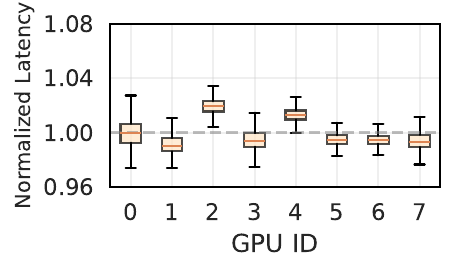}
        \caption{\mithreehundred system.}
        \label{fig:variability_mi300x_deepseek_prefill}
    \end{subfigure}
    \hfill
    \begin{subfigure}[t]{0.46\linewidth}
        \centering
        \includegraphics[width=\linewidth]{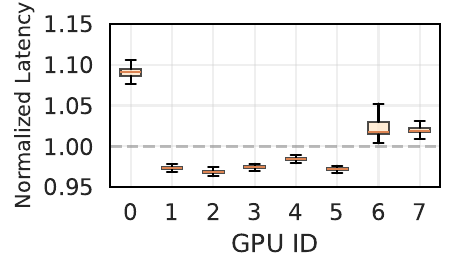}
        \caption{Skewed \mithreetwentyfive system.}
        \label{fig:variability_mi325x_deepseek_prefill_skewed}
    \end{subfigure}
    
    \caption{\seokjin{Kernel execution time variability for an MoE layer of DeepSeek-V3 under perfect token load balance. Latency is normalized by average across 8 GPUs for a single layer. }}
    \label{fig:other_variability}
\end{figure}

\begin{figure}[t]
    \centering
    \begin{subfigure}[b]{\linewidth}
        \centering
         \includegraphics[width=\linewidth]{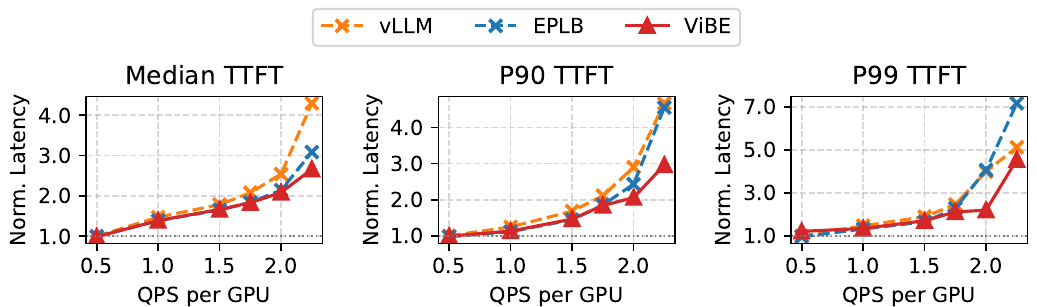}

        \caption{\seokjin{\mithreehundred system.}}
        \label{fig:mi300x_deepseek}
    \end{subfigure}
    
    \begin{subfigure}[b]{\linewidth}
        \centering
        \includegraphics[width=\linewidth]{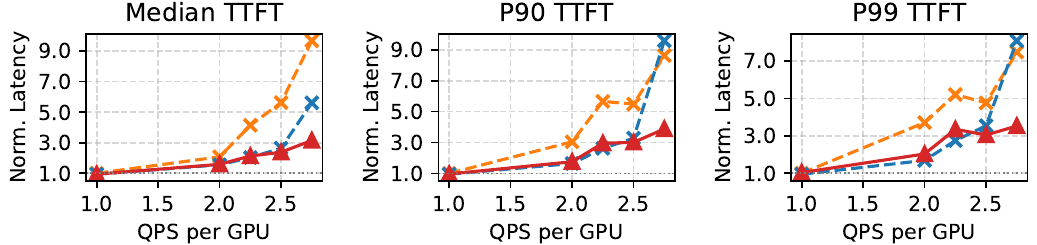}

        \caption{\seokjin{Skewed \mithreetwentyfive system.}}
        \label{fig:mi325x_deepseek_skewed}
    \end{subfigure}

    \caption{\seokjin{Key performance results of DeepSeek-V3 Sonnet prefill on different systems.}}
    \label{fig:mi325x_deepseek_skewed}
\end{figure}

\niparagraph{Sensitivity to Variability Distribution.}
Since process variation differs across every system, \ours measures each GPU individually and adapts to whatever distribution is present.
We validate across two regimes: (a)~an \mithreehundred node~\cite{mi300x} with lower variability, and (b)~a skewed scenario where we modify the voltage-frequency curve of GPU~0 in the \mithreetwentyfive system to induce up to 13\% performance deviation
(Figure~\ref{fig:other_variability}). We assign 4K tokens per expert, and report latency distribution across 1000 consecutive MoE layer executions.
Under mild variability (a), \ours provides consistent TTFT improvement over vLLM and EPLB across all QPS (Figure~\ref{fig:mi300x_deepseek}), confirming benefits even when the spread is modest.
Under skewed variability (b), \ours assigns fewer tokens to the degraded GPU, producing a more asymmetric distribution than EPLB.
At high QPS, where throttling activates, this directly reduces straggler latency and widens the gap over EPLB (Figure~\ref{fig:mi325x_deepseek_skewed}).

\begin{figure}[t]
    \centering
    \includegraphics[width=1\linewidth]{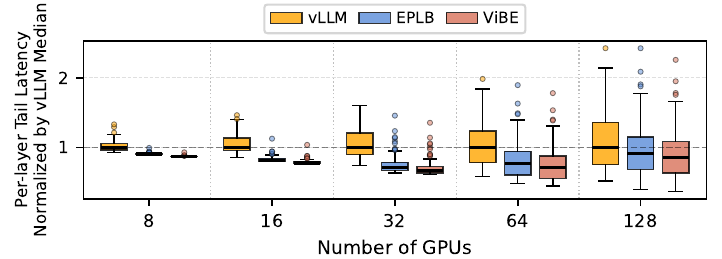}
    \caption{\seokjin{Projected per-MoE-layer tail latency versus EP group size, using measured 80$\times$ \mithreehundred performance profiles. 
    }}
    \label{fig:rackscale_projection}
\end{figure}

\niparagraph{Extending to Rack-Scale Systems.}
\seokjin{
Emerging rack-scale systems integrate over 70 GPUs in a single scale-up domain~\cite{helios,nvl72}.
We project placement behavior at larger EP group sizes using measured per-GPU profiles from 10 $8{\times}$\mithreehundred nodes~\cite{amduniv}, preserving the empirical variability
distribution while scaling group size. For larger GPU counts, we emulate the distribution by duplicating a subset of GPUs.
For each configuration, we run all three policies on ShareGPT token counts and project per-layer tail latency.
Figure~\ref{fig:rackscale_projection} shows that \ours's advantage depends on two competing factors.
Larger EP groups accumulate more performance spread, increasing straggler probability and the benefit of routing around slow devices.
However, each GPU holds fewer experts as the group grows, reducing placement flexibility.
These factors create a sweet spot at 16--32 GPUs; beyond 64 GPUs, all algorithms converge to nearly identical assignments as the per-GPU expert count collapses.
This motivates co-design strategies at extreme EP degrees---such as variability-aware TP grouping or selective expert duplication---which we leave to future work.
}

\section{Conclusion}

This work revisits a basic assumption in distributed MoE serving: that balancing token counts is sufficient to balance performance. We show that this assumption breaks down in practice because execution time is shaped not only by routing-induced workload skew, but also by GPU-to-GPU performance variability. Under expert parallelism, where each MoE layer completes at the pace of the slowest device, these effects combine to create persistent stragglers that inflate tail latency, reduce utilization, and limit sustainable throughput.
To address this, we propose \ours, a variability-informed expert placement framework that shifts the optimization target from token balance to execution-time balance. \ours combines per-GPU performance modeling, expert activation profiling, execution-time-aware placement, and drift-aware recalibration to align per-layer completion times across GPUs. By assigning highly-loaded experts to faster devices and less-loaded experts to slower ones, \ours uses variability as a lever to reduce stragglers rather than treating it as a source of inefficiency.
Across multiple MoE models, \ours consistently reduces execution-time imbalance, lowers tail latency, and expands the SLO frontier without sacrificing peak throughput. These results show that hardware variability is not a secondary artifact of deployment, but a first-order constraint in large-scale LLM serving. 
\section{Acknowledgements}

Claude Code was utilized to generate portions of this work, specifically for the creation of tables and figures, LaTeX formatting assistance, and assistance with ViBE code development.


\bibliographystyle{ACM-Reference-Format}
\bibliography{reference}
\clearpage
\newpage
\appendix
\section{Appendix}

\subsection{\ours End-to-End Flow}
\label{app:e2e}

Algorithm~\ref{alg:vibe} summarizes the complete \ours pipeline across its three phases.

\niparagraph{Phase 1: Offline profiling.}
Each GPU is profiled independently using the fused MoE kernel across a range of token counts, producing a device-specific latency model $f_g(n)$. This profiling is performed once per GPU and model configuration, as the relationship between token load and kernel latency is stable over time. In parallel, a representative workload $\mathcal{D}$ is run through the model to collect the per-layer per-expert activation matrix $\mathbf{W}$, capturing the routing skew that the placement solver must account for.

\niparagraph{Phase 2: Initial placement.}
For each MoE layer, \ours computes speed-proportional token targets $\tau_g$ from the profiled latency models and assigns experts greedily in descending load order to the GPU with the most remaining capacity relative to its target. This produces an initial placement that equalizes predicted execution time across GPUs, accounting for both workload skew and hardware variability.

\niparagraph{Phase 3: Online recalibration.}
During serving, \ours periodically compares the current expert load distribution against the reference snapshot recorded at the last rearrangement. When the cosine distance exceeds $\delta_{\cos}$ at any layer, indicating that the routing pattern has shifted, the system collects fresh routing statistics under the new distribution and triggers the incremental solver (Algorithm~\ref{alg:incremental}) to adjust the placement with minimal expert movement. After rearrangement, the reference snapshot is updated and a cooldown period suppresses spurious re-triggers from transient load bursts caused by the rearrangement itself.

\subsection{Incremental Placement Update}
\label{app:incremental}

Rather than recomputing placement from scratch upon workload drift, Algorithm~\ref{alg:incremental} iteratively swaps experts between the slowest and fastest GPUs, scoring each swap by its marginal latency reduction via $f_g$. For each layer, the solver identifies the GPU pair with the largest latency gap, evaluates all valid swaps between them, and applies the one with the greatest gain. The process stops once tail latency is within $(1{+}\epsilon)$ of the mean or no beneficial swap exists, typically converging in 5--30 swaps per layer versus over 200 reassignments for a full re-solve.

\SetKwFor{Every}{every}{do}{end every}

\FloatBarrier
\begin{algorithm}[t]
\small
\caption{\ours End-to-End Flow}
\label{alg:vibe}
\KwIn{Model with $L$ MoE layers, $E$ experts, GPU set $G$,
      workload $\mathcal{D}$, drift threshold $\delta_{\cos}$,
      tolerance $\epsilon$, monitoring interval $H$}
\KwOut{Expert placement $\mathcal{S} = \{S_g^{(l)}\}$}

\tcc{Phase 1: Offline profiling}
\ForEach{GPU $g \in G$}{
    Profile MoE kernel $\rightarrow f_g(n)$\;
}
Run $\mathcal{D}$ through Model $\rightarrow \mathbf{W} \in \mathbb{R}^{L \times E}$\;

\tcc{Phase 2: Initial placement (per layer)}
\ForEach{layer $l = 1 \ldots L$}{
    $N \gets \sum_e w_e^{(l)}$\;
    $s_g \gets 1 / f_g(N / E)$ for each $g$\;
    $\tau_g \gets N \cdot s_g / \sum_h s_h$\;
    Sort experts by $w_e^{(l)}$ descending\;
    \ForEach{expert $e$ in sorted order}{
        Assign $e$ to $\arg\max_g (\tau_g - n_g)$\;
        $n_g \gets n_g + w_e^{(l)}$\;
    }
}

\tcc{Phase 3: Online serving with recalibration}
Snapshot $\hat{\mathbf{w}}^{(l)} \gets$ current load vectors\;
\While{serving}{
    \Every{$H$ forward passes}{
        \ForEach{layer $l$}{
            $d \gets 1 - \frac{\mathbf{w}^{(l)} \cdot \hat{\mathbf{w}}^{(l)}}{\|\mathbf{w}^{(l)}\|\;\|\hat{\mathbf{w}}^{(l)}\|}$\;
        }
        
        \If{$\max_l d > \delta_{\cos}$}{
            Update $\mathbf{W}$ from recent routing\;
            IncrementalUpdate($\mathcal{S}$, $\mathbf{W}$, $\{f_g\}$, $\epsilon$)\;
            $\hat{\mathbf{w}}^{(l)} \gets \mathbf{w}^{(l)}$ for all $l$\;
            Cooldown for $H$ forward passes\;
        }

    }
}
\end{algorithm}

\FloatBarrier
\begin{algorithm}[t]
\small
\caption{Incremental Placement Update}
\label{alg:incremental}
\KwIn{Placement $\mathcal{S}$, expert token load $\{w_e\}$,
      latency models $\{f_g\}$, tolerance $\epsilon$}
\KwOut{Updated placement $\mathcal{S}'$ with minimal expert moves}
\ForEach{layer $l$}{
    Compute per-GPU token load $n_g \gets \sum_{e \in S_g} w_e$\;
    Compute target latency $\bar{f} \gets \frac{1}{|G|}\sum_g f_g(n_g)$\;
    \Repeat{
        $\max_g f_g(n_g) \leq (1{+}\epsilon)\,\bar{f}$
        \textbf{ or } no latency reduction
    }{
        $g_{+} \gets$ GPU with highest $f_g(n_g)$
            \tcp*{slowest GPU}
        $g_{-} \gets$ GPU with lowest $f_g(n_g)$
            \tcp*{fastest GPU}
        Find best swap $(e_i, e_j)$ between $g_{+}$ and $g_{-}$
            maximizing latency reduction\;
        Swap $e_i \leftrightarrow e_j$\;
        Update $n_{g_{+}},\; n_{g_{-}},\; \bar{f}$\;
    }
}
\end{algorithm}


\end{document}